# Design principles of natural light harvesting as revealed by single molecule spectroscopy

T.P.J. Krüger[a,*], R. van Grondelle[b]

[a]*Department of Physics, University of Pretoria, Private bag X20, Hatfield 0028, South Africa*
[b]*Department of Physics and Astronomy, VU University Amsterdam, De Boelelaan 1081, 1081 HV Amsterdam, the Netherlands*

*Corresponding author. Tel.: +27124202508. E-mail address: tjaart.kruger@up.ac.za (T.P.J. Krüger)

Biology offers a boundless source of adaptation, innovation, and inspiration. A wide range of photosynthetic organisms exist that are capable of harvesting solar light in an exceptionally efficient way, using abundant and low-cost materials. These natural light-harvesting complexes consist of proteins that strongly bind a high density of chromophores to capture solar photons and rapidly transfer the excitation energy to the photochemical reaction centre. The amount of harvested light is also delicately tuned to the level of solar radiation to maintain a constant energy throughput at the reaction centre and avoid the accumulation of the products of charge separation. In this Review, recent developments in the understanding of light harvesting by plants will be discussed, based on results obtained from single molecule spectroscopy studies. Three design principles of the main light-harvesting antenna of plants will be highlighted: (a) fine, photoactive control over the intrinsic protein disorder to efficiently use intrinsically available thermal energy dissipation mechanisms; (b) the design of the protein microenvironment of a low-energy chromophore dimer to control the amount of shade absorption; (c) the design of the exciton manifold to ensure efficient funneling of the harvested light to the terminal emitter cluster.

# 1. Introduction

Artificial photosynthesis is envisioned by many to be an important component of mankind's long-term energy solution [1]. Bioinspired photosystems appear most promising, but the first constructs over the past few years have clearly pointed to the infancy of this field [2-4]. To make progress, a very detailed understanding of natural photosynthesis is required in order to wisely extract the most important design principles. Here, the primary steps of photosynthesis – light harvesting and charge separation – are the most crucial to ensure that the energy of an absorbed photon is stored with a sufficiently high probability, which is commonly 90-100% under conditions of low solar radiation! The design principles of charge separation, which takes place in the so-called reaction centre, are now beginning to be understood. The speed and efficiency of charge separation are based on a finely designed structure that minimises free energy losses, enables selected vibrations to drive quantum coherent processes, and allows control over the multiple pathways that can be followed by an excitation in the reaction centre [5]. The process of photosynthetic light harvesting has proven to be even more complex. Although a few important design principles can be identified for the purpose of designing synthetic systems [6], many mechanistic details are still incomplete, and further experimental and theoretical advances are awaited to deepen our understanding. One such promising technique is known as single molecule spectroscopy (SMS) and will be the main focus of this Review.

Photosynthetic light harvesting is performed by an array of interacting chromophores that absorb (solar) photons and transfer the resulting electronic excitation energy to the reaction centre. The chromophores are typically held in fixed positions and orientations by proteins; yet, the protein is much more than a scaffold. It also interacts strongly with the chromophores, thereby significantly altering their spectroscopic and light-harvesting properties. The unique properties of the protein, which underlie its interaction with the chromophores, are unmatched in any solar energy technological device: (a) the protein constitutes a highly heterogeneous dielectric environment, which provides every chromophore with a unique transition energy (also referred to as "site energy") and strongly modifies the electronic couplings amongst the chromophores; (b) the protein is a highly dynamic structure, exhibiting motions on timescales ranging from sub-ps to >1 s, a behaviour commonly referred to as disorder. The structural disorder is not only translated into time-dependent fluctuations of the site energies, but also gives rise to phonons – fast, collective nuclear vibrational modes of the protein – which interact with the electronic excited states of the chromophores and consequently change the energy-transfer dynamics.

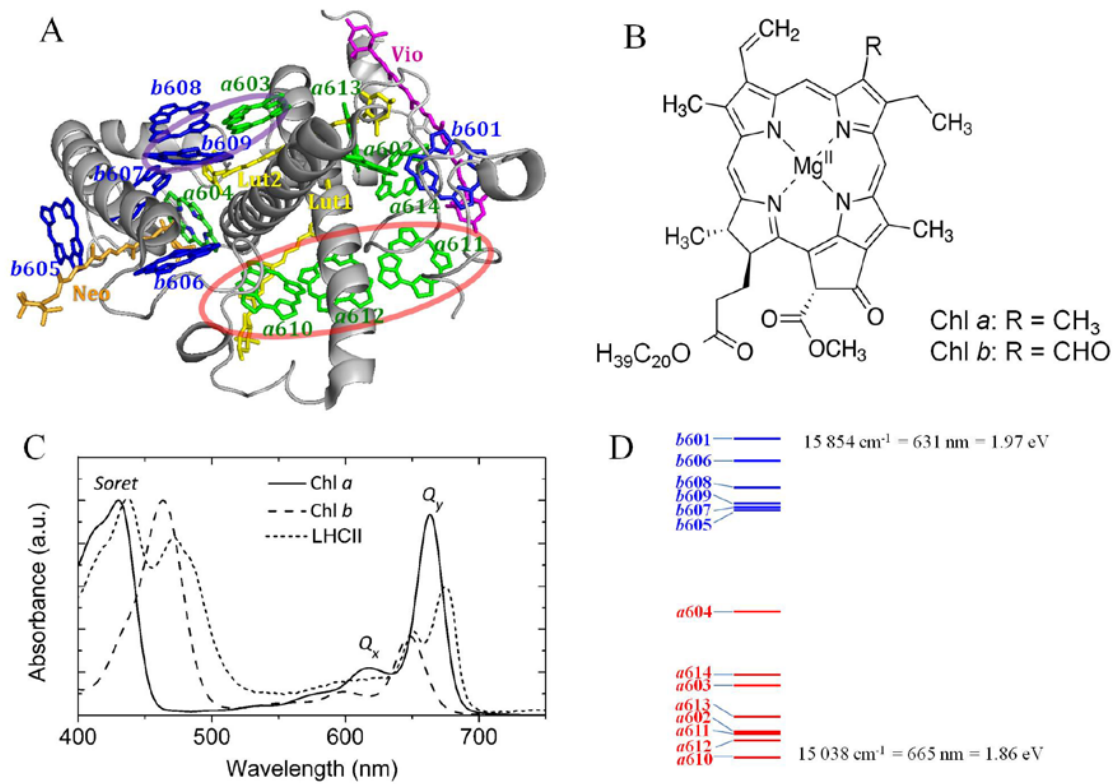

**Fig. 1:** (A) 2.72 Å crystal structure according to Ref. [8], shown from the outside of the membrane (*i.e.*, stromal view), and using the authors' nomenclature. For clarity, only the chlorin rings of Chl *a* and *b* are shown, in gray and black (online green and blue), respectively. The protein is displayed as a gray ribbon and the carotenoids lutein, neoxanthin, and violaxanthin are denoted by Lut, Neo, and Vio, respectively. Encircled are the two Chl clusters that are discussed in the text, *viz.*, 603–609 and 610–611–612. (B) Chemical structure of Chl, differing only at the residue, R, defined bottom right. (C) Room-temperature absorption spectrum of Chl *a* (*solid line*) and Chl *b* (*dashed line*) in ethanol, as well as LHC2 (*dotted line*). All three spectra are normalised to the maximum (*i.e.*, the Soret peak). The three dominant absorption bands, Soret, $Q_x$ and $Q_y$, are indicated for Chl *a*. (D) Relative values of all Chl site energies in each subunit of LHC2, according to the model for the trimer in Ref. [16]. Relative energies levels of Chls *a* (bottom) and *b* (top). The values of the highest and lowest energies are indicated.

Photosynthetic chromophores are remarkable molecules in several respects. Consider the main light-harvesting complex (LHC) of plants, LHC2, which naturally assembles into a three-fold symmetric structure (*i.e.*, trimer) of identical subunits (*i.e.*, monomers) (**Fig. 1A**), each containing no less than 14 chlorophylls (Chls) [7, 8]. The large conjugated ring of Chl supplies the molecule with a substantial absorption cross section as well as a rigid structure that cannot be easily deformed. Yet, a relatively small modification of the ring can dramatically shift the transition energies: the two types of Chls found in LHC2 – Chl *a* and *b* – differ only at the small side chain *R* (**Fig. 1B**), but this structural change leads to a 30 nm (*i.e.,* 0.085 eV) shift of the lowest electronic transition (*i.e.*, HOMO to LUMO, also known as $Q_y$) (**Fig. 1C**). Likewise, the transition energies of the embedded Chls are tuned by the protein matrix across a large range, providing each Chl in LHC2 with a unique site energy (**Fig. 1D**). This not only significantly increases the absorption spectral window but also creates an energy gradient, so that the excitation energy can be "funnelled" to a particular site in the LHC or to the reaction centre [9, 10]. Another notable property of Chls is their electronic excited states having intrinsic decay times of a few ns. This is six–seven orders of magnitude longer than the timescale of absorption and two–three orders of magnitude longer than the timescale of energy transfer to the reaction centre. In the natural environment, spontaneous emission is therefore a negligible decay channel of the excitation and sufficient time is allowed to initiate the first steps of charge separation. When an LHC is isolated from its natural environment, the fluorescence yield obviously increases considerably and can be used as a probe. Finally, Chls are not only used by plant photosystems as efficient light harvesters but also as energy sinks within an antenna complex or the reaction centre. To this end, a Chl–Chl pair is used to create a charge-transfer state, which can rapidly deplete excitation energy in the antenna or initiate the process of charge separation in the reaction centre [11]. Each monomeric subunit of LHC2 binds four additional chromophores, known as carotenoids (see Fig. 1A). These molecules extend the absorption spectral window by harvesting solar energy in the blue–green spectral region, which is then rapidly transferred to the Chls [12]. Even more important is their photoprotective role whereby Chl singlet and triplet states are efficiently quenched: Chl triplets would otherwise react with oxygen to produce highly reactive (and therefore lethal) singlet oxygen [13]; Chl singlets are quenched when the excitation rate of the photosystem becomes too high (*vide infra*) [14, 15].

The chromophores in LHCs occur at an astounding density. For example, the Chl concentration of LHC2 is 0.25 M, which gives Chl–Chl separations as short as 9–10 Å and strong (up to 110 cm$^{-1}$ [16]) interactions amongst the Chls. When the Chls are solubilised at this concentration in an organic solvent with the same average dielectric as the protein, the fluorescence will be virtually zero due to a process known as concentration quenching [17]. Although the arrangement of Chls in LHC2 may appear random, they are actually perfect for optimisation of the energy transfer [6, 10, 18]. One important reason is that the high chromophore density creates new physical states, known as excitons (the specific type being Frenkel excitons or molecular excitons), whereby the excitation is delocalised over a number of chromophores and hence coherently shared [19]. Excitons significantly decrease the number of pathways that need to be explored during energy migration to the reaction centre, thus leading to shorter transfer times and larger quantum efficiencies. Excitation traps due to single site defects in the antenna network can also be avoided more easily by such a delocalised excitation. The exciton delocalisation length in LHCs is typically 2–4 chromophores, which conforms to the model of LHC2 consisting of strong excitonically coupled clusters of 2–4 Chls [10]. In LHC2, the cluster Chl *a*610-*a*611-*a*612 contributes most strongly to the lowest three exciton states and therefore constitutes the site where the excitation in LHC2 will most likely end up. For this reason it is called the terminal emitter cluster. Not surprisingly, this cluster also neighbours other LHCs in the antenna network of Photosystem 2 of plants and is therefore the preferred terminal site of LHC2 [20, 21].

If organic components are considered for solar energy technologies, the necessary photoprotective measures have to be taken, because biological systems are prone to photodamage. For example, the protection provided by carotenoids is not sufficient for LHC2. A complex set of mechanisms is used to account for gradual and rapid changes in solar intensity during the day and over longer periods (*e.g.*, seasonal changes). This level of regulation is a far cry from what any artificial solar cell is capable of. An important part of this self-protecting feedback regulation takes place in the antenna network of Photosystem 2 and can be observed as non-photochemical quenching (NPQ) of Chl *a* fluorescence [22]. LHC2 is prominently involved with the fastest, rapidly reversible component of NPQ by thermally dissipating excess absorbed energy before it migrates further in the photosystem [22, 23]. The main molecular mechanism was identified as a protein structural change which greatly increases the probability of energy transfer from the terminal emitter Chl cluster to the lowest electronic excitation energy level (LUMO or $S_1$) of the nearby carotenoid (Lut 1), the latter state of which relaxes quickly and nonradiatively [14]. The same mechanism, involving Chl *a* and *β*-carotene, was very recently identified in a high light-inducible cyanobacterial protein complex [15].

What can we expect SMS to contribute to our understanding of photosynthetic light harvesting? The strength of SMS lies in its sensitivity and selectivity by resolving spectroscopic dynamics that are otherwise hidden in the ensemble average. Traditional SMS techniques typically resolve fluorescence fluctuations on timescales spanning ms to a few mins [24], thus providing a unique lens on the static disorder, *i.e.*, relatively slow protein structural fluctuations. Noteworthy is that these fluctuations generally correspond to relatively large motions of protein subdomains, which are often related to functional changes. In addition, spectroscopic heterogeneities can be identified and divided between complex-to-complex variations and time-dependent fluctuations of a single complex. A full statistical description of spectroscopic observables can thus be obtained instead of a single mean value. Finally, rare photophysical events may be discovered, which may not play an obvious biological functional role, but nevertheless lead to a more complete understanding of the energy landscape and the photophysical and photobiological behaviour of LHCs. In this Review, three design principles in addition to those mentioned above will be pointed out, based on recent resuls from SMS experiments and modelling on LHC2 and related plant LHCs.

## 2. Three Design Principles of Natural Light Harvesting Discovered With SMS

In order to make most of an SMS measurement, one needs to find a balance between the signal-to-noise ratio (SNR) of the measured signal and the survival time of the complex. A higher excitation intensity obviously increases the former but it also decreases the latter. Still, the SNR often does not increase indefinitely but after some intensity threshold starts to decrease again due to the presence of long-living triplet states, which annihilate any subsequent excitations in the complex, leading to fluorescence saturation. For LHC2, the optimal SNR corresponds to a typical excitation rate of ~$10^6$ s$^{-1}$ [25, 26]. The SNR is furthermore optimised by ensuring minimal loss in the detection branch of the experimental setup and employing very sensitive detectors. In addition, background noise, arising mainly from Raman scattering of water, is limited by using a diffraction-limited excitation volume and a confocal pinhole. An optimal detection time per complex is ensured by immobilising the complex, typically on a flat surface via a soft, electrostatic interaction. A raster-scanned fluorescence image across an identified surface area exposes the position of every immobilised complex (**Fig. 2A**). The excitation light is subsequently focussed on one complex of interest at a time. Various spectroscopic properties can then be probed, amongst the most common being fluorescence spectral diffusion (**Fig. 2B**) and fluorescence intensity fluctuations (**Fig. 2C**).

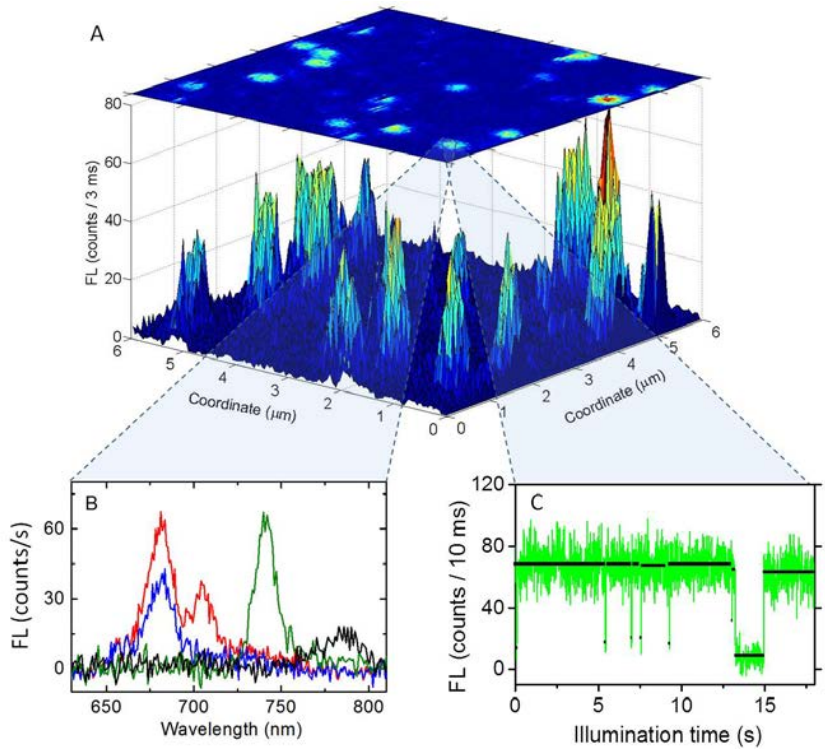

**Fig. 2**. (A) Example of a raster-scanned fluorescence (FL) map of a few surface-bound LHC2 trimers. (B) Different spectral shapes observed from individual complexes, each of which was reversibly accessed and quasi-stable for periods of up to minutes. An acquisition time of 1 s was used. (C) Rapidly fluctuating wavelength-integrated FL intensities (online green) and resolved intensity levels (black) using the algorithm described in Ref. [51].

## 2.1. Photoactive control over the intrinsic protein disorder

The large, rapid, reversible intensity changes shown in **Fig. 2C** is a phenomenon known as fluorescence intermittency or blinking and is one of the most evident examples of hidden information in ensemble-averaging measurements. Curiously, almost every nanoscale fluorescent object is known to exhibit fluorescence blinking. For semiconductor quantum dots and solubilised chromophores, the dominant underlying molecular mechanism was shown to involve ionisation, in the former due to a photoassisted Auger process [27, 28] and in the latter, long-living, dark radical states were populated via triplet states [29]. For LHC2, containing 54 chromophores, one might similarly expect radical states to be responsible for fluorescence blinking. However, to date there is no experimental evidence for this hypothesis. So far, radicals have only been reported from LHC2 carotenoids after very specific double excitation using resonant two color, two photon ionisation spectroscopy [30], though these radicals existed for sub-nanoseconds. Considering the large amount of singlet–triplet annihilation in a typical SMS experiment, one might consider the possibility of radical formation after excited state absorption of a triplet state. This possibility was investigated in two multipulse experiments (TPJ Krüger and B van Oort, unpublished results). In the first experiment, the light conditions of SMS experiments were reproduced by using a pump pulse and a repump pulse, 1 µs apart and both at 630 nm. In the second experiment, carotenoid triplets were excited to higher energy levels by first pumping the complexes at 630 nm and later at 510 nm after complete relaxation of the singlet excitation state. The data obtained from probing the transient absorption states across the range of 450 – 970 nm showed no evidence of radical state formation. We conclude that the timescale of relaxation of a dark state in fluorescence blinking data (*i.e.*, ms to tens of seconds) is much longer than all other photophysical events and rather points to static protein disorder, which changes the probability of the excitation to access one or more traps in the complex.

To shed more light on the nature of the long-living dark states of LHC2, fluorescence blinking was investigated in the context of NPQ for all the peripheral light-harvesting complexes of plant Photosystem 2, *i.e.*, LHC2 and the so-called monomeric minor complexes [31, 32]. To this end, different conditions known to be involved with NPQ in the natural environment were mimicked in the SMS setup. Specifically, an acidic environment was used, the concentration of the detergent, which mimics the membrane within which the complexes are naturally embedded, was drastically decreased, and one of the carotenoids, violaxanthin, was replaced with zeaxanthin. The combination of the environmental changes gave rise to a significantly increased average dwell time in quenched states for LHC2 (**Fig. 3A**), while the opposite was observed for the minor antenna complexes (**Fig. 3B**). Moreover, it was also shown that the switch into the dark states of LHC2 is strongly light-induced [33]. It can be concluded that the molecular mechanism underlying fluorescence blinking in LHC2 plays an important role in NPQ. In other words, an intrinsically available thermal energy dissipation state is used for the purpose of photoprotection and the probability of accessing this state is strongly enhanced under NPQ conditions. The large intensity fluctuations from a single LHC2 complex (**Fig. 2C**) reflect the static disorder. Switching between mainly a "light" and a "dark" state corresponds to high and low probabilities of accessing the quenched state, respectively. This idea was used to model fluorescence blinking in LHC2, based on structural diffusion of the protein surrounding the terminal emitter Chl domain, which would determine the probability of excitation energy transfer to the $S_1$ state of Lut 1 according to one of the NPQ mechanisms [34, 35]. By considering static disorder to modulate access to the dark state, most of the intensity dwell time statistics of the experimental SMS data was reproduced [34, 35]. A future extension of this model could seek to reproduce the quasi-continuum of stable fluorescence levels (**Fig. 2C**), which points to numerous quasi-stable structural states of the protein.

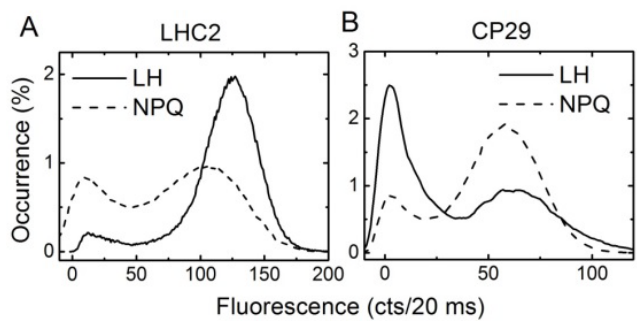

**Fig. 3**: Intensity distributions of an ensemble of individually measured LHC2 trimers (A) and CP29 (B), the latter of which is representative of the minor antenna complexes. Distributions are displayed for an environment mimicking low light (light harvesting, LH, *solid lines*) and high light (quenching, NPQ, *dashed lines*) conditions (see Refs. [31, 32]).

From the preceding discussion it can be concluded that the $S_1$ state of Lut 1 is one important photoprotective energy trap in LHC2. A very recent SMS study revealed the presence of an additional energy dissipative state in LHC2, which was found to be accessed more frequently in an acidic environment and for a zeaxanthin-enriched mutant, both of which are NPQ-related conditions [36]. Furthermore, Stark fluorescence experiments have indicated that excitation-dissipating charge-transfer states appear when LHC2 forms aggregates, another state representing NPQ [37]. Recently, a multiphoton experimental study has disclosed yet another dark state in LHC2, which was suggested to be related to fluorescence blinking and which may also explain the strong quenching observed from LHC2 aggregates

[38]. All the above-mentioned dark states were suggested to be accessed after a protein conformational change. It can be concluded from the SMS results that photoprotective energy dissipation in LHC2 is determined by fine environmental control over the protein disorder, determining the probability of accessing different possible excitation traps in the complex.

## 2.2. Control of shade light absorption

To shed more light on the nature of the spectral diffusion displayed by LHC2 (**Fig. 2B**), a disordered exciton model, based on modified Redfield theory, was used [16, 26]. The static disorder was simulated by randomly varying each of the site energies of the Chls around their average values and calculating the resulting energy equilibration and fluorescence spectral shape. A small fraction of realisations of the disorder (1–3%, depending on the width of the disorder assumed in the model) gave rise to spectral shapes that deviate from the ensemble-averaged spectrum. These deviating spectral shapes had a maximum intensity that peaked within ~10 nm of the ensemble spectral peak at ~682 nm and corresponded well to the experimentally measured single molecule spectra. All these relatively small spectral variations can therefore be explained by slow protein structural disorder, which induces variations of the site energies, which in turn determine the specific pattern of the exciton state before it relaxes via fluorescence.

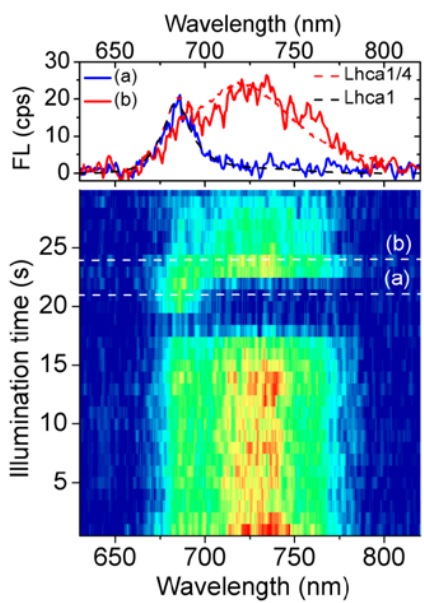

**Fig. 4**: Representative example of time-resolved fluorescence spectra from a single Lhca1/4 dimer that switched off its entire red spectral component (a) and completely recovered (b) [47]. Spectra on top represent the spectra at the white, horizontal, dashed lines. Black, dashed spectrum and gray (online: red), dashed spectra represent a bulk spectrum of an Lhca1 monomer and an Lhca1/4 dimer, respectively.

However, site energy disorder fails to explain various other spectra exhibited by LHC2: ~3–5% of the complexes reversibly visited spectral states with a peak wavelength at ~700 nm [26, 39], while numerous spectra have been observed at even longer wavelengths, occasionally with shifts of more than 100 nm to the red [32]! These red-shifted states cannot be explained by site-energy disorder but require the protein to switch into particular conformations to invoke special chromophore interactions. How unusual are these low-energy states? First, emission at 700 nm is widely considered to be a signature of NPQ [40, 41]. Second, two of the peripheral antenna complexes of plant Photosystem 1 are typified by fluorescence spectral peaks beyond 720 nm, despite their proteins and chromophore content being remarkably similar to that of LHC2 [42]. The red emission from these Photosystem 1 antennae originates from a Chl dimer (Chls 603 and 609) that is so closely spaced that a charge-transfer state forms, which mixes with the lowest excitonic states, giving rise to a considerable increase in the reorganisation energy [43]. This process was experimentally observed by the strong effect of an externally applied electrical field on the red absorption/emission of the Chl dimer [44]. In LHC2, the distance between Chls 603 and 609 is slightly larger, mainly due the specific protein microenvironment: Chl 603 in LHC2 binds a histidine instead of an asparagine [8, 45]. The special protein conformations that give rise to ≥700 nm emission from LHC2 most likely decrease the distance between Chls 603 and 609, thereby invoking an exciton–charge-transfer mixed state [46]. SMS measurements on Photosystem 1 antenna complexes have demonstrated that these complexes are capable of switching off the charge-transfer character of the Chl 603–609 dimer (**Fig. 4**) [47]. Only a subtle protein structural change is necessary to achieve this spectral tuning [48] and hence may dramatically alter the capability of the complexes to absorb low-energy photons. The functional significance of this behaviour can be appreciated by considering that almost all visible light is absorbed by a single leaf, so that shade light contains an exceptionally high ratio of near-infrared (*i.e.*, >700 nm) photons. Curiously, the appearance of 700 nm emission under NPQ conditions suggests that the photoprotective state actually enhances shade light absorption. We conclude that the amount of shade light absorption by Photosystems 1 and 2 is controlled by the specific protein microenvironment of the Chl 603–609 dimer through subtle structural changes brought about by the nature of a single amino acid or modulated by NPQ conditions.

## 2.3. Exciton delocalisation ensures fast and robust light harvesting

How crucial is the terminal emitter Chl cluster of LHC2 for its light-harvesting function? Stated more generally, how would the quantum efficiency of Photosystem 2 be affected if the delocalised exciton character of the terminal emitter Chl cluster in LHC2 is diminished? These questions were recently addressed in an SMS study of an LHC2 mutant, known as LHC2-A2, in which the Chls $a$611 and $a$612 had been knocked out [49]. The strong excitonically coupled cluster of $a$610-$a$611-$a$612 was hence reduced to only Chl $a$610. This mutation gave rise to some prominent spectral changes, which include considerable heterogeneity of the fluorescence spectra. In particular, ~50% of the complexes exhibited stable emission in a narrow band centred at 678 nm (*i.e.*, 3–4 nm blue-shifted with respect to emission from wild type LHC2), while the emission from the other ~50% was spread in a stable band broadened towards the red, giving rise to a peak maximum up to ~684 nm. A fraction of complexes (~5%) were observed to switch between the two types of emission states (Fig. 5), pointing to a protein conformational change. A modified Redfield – disordered exciton model was again used to shed light on the underlying excitonic pattern. Localisation of the energy on the single Chl $a$610 led to a substantial increase in the reorganisation energy and was responsible for an emission band at ~688 nm, while the 678-nm emission originated mainly from the strong excitonically coupled Chl $a$602-$a$603 cluster. The broad spectra (**Fig. 5, gray, online green**) thus correspond to conformational states where emission occurred from both $a$610 and $a$602-$a$603.

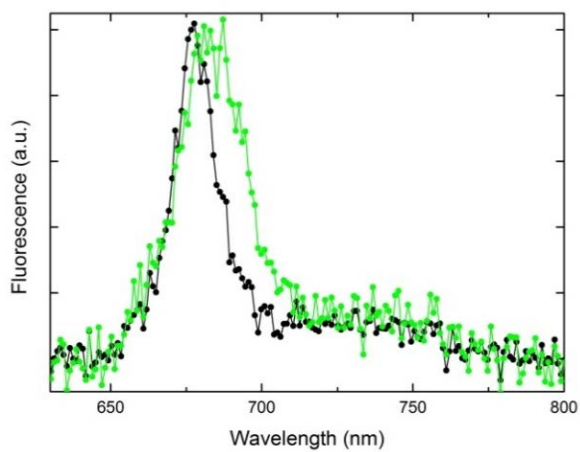

**Fig. 5** Two fluorescence spectral shapes between which a single monomer of the LHC2-A2 mutant switched [49].

What do the above-mentioned results tell us about the light-harvesting efficiency of LHC2-A2? First, the experimental results (*i.e.*, enhanced spectral heterogeneity) point to an excitonic manifold that is more sensitive to static disorder because of the significantly reduced exciton delocalisation at the terminal emitter site. Second, the modelling results show that the three lowest exciton states of wild-type LHC2 are separated by only 30 cm$^{-1}$ and involve mostly participation of $a$610-$a$611-$a$612, so that the excitation will end up on the terminal emitter cluster for basically all realisations of the disorder and can do so via numerous paths. In contrast, in the mutant, Chl $a$602 contributes strongly to the lowest exciton state, which means that a substantial fraction of excitations end up at this unfavourable site, located on the inside of the trimer [8]. This slows down transport to neighbouring antenna complexes and to the reaction centre. In addition, the mutant complex was often found in a conformation where the excitation was almost completely trapped at $a$602-$a$603 (see **Fig. 5,** *black spectrum*), pointing to the complex's large sensitivity to static disorder. The average energy transfer rate from Chl $a$602 to $a$610 was, furthermore, 35% slower in the mutant than in the wild-type complex thanks to fewer energy pathways to Chl $a$610, reducing the mutant's light-harvesting efficiency even further. Very recent quasi-elastic neutron scattering results of the same mutant confirmed the robustness of the terminal emitter cluster to protein disorder [50].

## 3. Conclusion

In conclusion, there are numerous design principles that collectively bring about a remarkable light-harvesting efficiency of the main light-harvesting complex of plants, LHC2, as well as a staggering level of adaptability to environmental influences, especially varying levels of solar radiation. In addition to the design motifs that have been known for a few decades, such as the special character of the light-harvesting chromophores, their high density, the protein host, and the molecular excitons, SMS has shed light on a few more principles. Firstly, a large degree of photoprotection is established by LHC2 using intrinsically available thermal energy dissipation channels by finely controlling the structural protein disorder on timescales of ms to tens of seconds. Secondly, based on the spectral dynamics of light-harvesting complexes from plant Photosystems 1 and 2, it was shown that the particular protein microenvironment of a Chl dimer is responsible for considerable tuning of the extent of shade absorption of plants. Finally, using a mutant of LHC2 where the terminal emitter chromophore cluster is disrupted, it was demonstrated that the strong exciton delocalisation of the terminal emitter cluster in wildtype LHC2 is responsible for energy transfer robustness despite the prevailing static disorder.

## Acknowledgements

The authors acknowledge past and present collaborators on this work: C. Ilioaia, R. Croce, E. Wientjes, C. Ramanan, J.M. Gruber, M. Negretti, M.P. Johnson, A.V. Ruban, P. Horton, and V.I. Novoderezhkin. T.P.J.K. was supported by the University of Pretoria's Research Development Programme (Grant No. A0W679). R.v.G. was supported by an Advanced Investigator grant from the European Research Council (No. 267333, PHOTPROT), Nederlandse Organisatie voor Wetenschappelijk Onderzoek, Council of Chemical Sciences (NWO-CW) via a TOP-grant (700.58.305), and by the EU FP7 project PAPETS (GA 323901). Financial assistance of the National Research Foundation (NRF), South Africa, is gratefully acknowledged. Any opinion, findings and conclusions or recommendations expressed in this article are those of the authors, and therefore the NRF does not accept liability in regards thereto.